\documentclass{llncs}
\usepackage{etex}
\usepackage{makeidx}  

\usepackage{amssymb}
\usepackage{amsmath}
\usepackage{amsfonts}
\usepackage[utf8]{inputenc} 
\usepackage{amsfonts}
\usepackage{pslatex}
\usepackage{subfigure}
\usepackage{graphicx}
\usepackage{listings}
\usepackage{color}
\usepackage{url}
\usepackage{paralist}
\usepackage{enumerate}
\usepackage[all,2cell,ps]{xy} 

\newbox\subfigbox
\makeatletter

\lstdefinelanguage{Algo}{ morekeywords={do,let,if,then,else,elseif,fi,while,endwhile,for,return,function,done,new,end,endif,and,or,switch,case,in,instanceof,deliver},
  sensitive, %
  morecomment=[l]\#,%
  morecomment=[l]//,%
  morecomment=[s]{/*}{*/},%
  morestring=[b]'',%
  morestring=[b]'%
}[keywords,comments,strings]

	{\def\caption##1{\gdef\subcapsave{\relax##1}}%
  		\let\subcapsave\@empty%
  		\setbox\subfigbox\hbox%
  		\bgroup%
	}%
	{%
		\egroup%
  	\subfigure[\subcapsave]{\box\subfigbox}%
  	}%

\makeatother

\definecolor{emph}{rgb}{0.45,0.45,0.45}
\definecolor{green}{rgb}{0,0.7,0}
\definecolor{grey}{rgb}{0.95,0.95,0.95} 

{\end{trivlist}}

\newcommand{\map}{\rightarrow}

\def\ID{ID}

\newcommand{\idComp}[2]{(#1:#2)}

\begin{document}

\title{Scalable XML Collaborative Editing with Undo\\
short paper}
\author{St\'ephane Martin\inst{1} \and Pascal Urso\inst{2} \and St\'ephane Weiss\inst{2}}
\institute{
\email{stephane.martin@lif.univ-mrs.fr}\\
Laboratoire d'Informatique Fondamentale\\
 39 rue F. Jolio-Curie,\\
 13013 Marseille, France
\and
\email{(pascal.urso,stephane.weiss)@loria.fr}\\
Universit\'e de Lorraine\\
LORIA, Campus Scientifique,\\  
54506 Vandoeuvre-l\`es-Nancy, France \\
}
\maketitle
\begin{abstract}

  Commutative Replicated Data-Type (CRDT) is a new class of algorithms
  that ensures scalable consistency of replicated data. It has been
  successfully applied to collaborative editing of texts without
  complex concurrency control.

  In this paper, we present a CRDT to edit XML data. Compared to
  existing approaches for XML collaborative editing, our approach is
  more scalable and handles all the XML editing aspects : elements,
  contents, attributes and undo. Indeed, undo is recognized as an
  important feature for collaborative editing that allows to overcome
  system complexity through error recovery or collaborative conflict
  resolution.
  
\end{abstract}

\section{Introduction}


In large-scale infrastructures such as clouds or peer-to-peer networks,
data are replicated to ensure availability, efficiency and
fault-tolerance. Since data are the heart of the information systems,
the consistency of the replicas is a key issue. Mechanisms to
ensure strong consistency levels -- such as linear or atomic -- do not
scale, thus modern large-scale infrastructures now rely on
eventual consistency.


Commutative Replicated Data Types~\cite{preguica09commutative,weiss09logoot} (CRDT)
is a promising new class of algorithms used to build operation-based
optimistic replication~\cite{saito05optimistic} mechanisms. It ensures
eventual consistency of replicated data without complex concurrency
control. It has been successfully applied to scalable collaborative
editing of textual document but
not yet on semi-structured data types.
%
%
EXtensible Markup Language (XML) is used in a wide range of
information systems from semi-structured data storing to
query. Moreover, XML is the standard format for exchanging data,
allowing interoperability and openness.


Collaborative editing (CE) provides
several advantages such as obtaining different viewpoints, reducing
task completion time, and obtaining a more accurate final result. Nowadays, 
collaborative editing becomes massive and part of our every
day life. The online encyclopedia Wikipedia users have produced 15
millions of articles in a few years. 

Undo has
been recognized as an important feature of single and collaborative
editors~\cite{abowd92undo,choudhary95general}. The undo feature
provides a powerful way to recover from errors and 
vandalism acts or to manage edit conflicts. 
So, it helps the user to face the complexity of the
system.
However, designing an undo feature is a non-trivial task. First, in
collaborative editing, this feature must allow to undo any operation --
and not only the last one -- from any
user. This is called global
selective undo (or anyundo). Second, this undo must be correct from
the user's point of view.  The system must return in a state such as
the undone operation has never been performed.






We propose to design an XML CRDT for collaborative editing. This CRDT
handles both aspects of XML trees : elements' children and
attributes. The order in the list of the elements' children are
treated as in linear structure CRDT. Elements' attributes are treated
using a last-writer-wins rule.
Our undo is obtained by keeping the previous value given to attributes
and operations applied on elements, and then counting concurrent undo
and redo operations. A garbage collection mechanism is presented to
garbage old operations.

%
 
\section{State of the art}
\label{sec:soa}

%


The Operational Transformation (OT) \cite{ressel96integrating}
approach is an operation-based replication mechanism. OT relies on a
generic integration algorithm and a set of transformation functions
specific to the type of replicated data. Some integration mechanism
use states vectors -- or context vectors~\cite{sun06operation} in
the presence of undo -- to detect concurrency between operations; such
mechanisms are not adapted to large-scale
infrastructures. Ignat et al. \cite{ignat08xml} propose to couples an integration
mechanism that uses anti-entropy, with some
specific transformation functions~\cite{oster06tombstone} to obtain P2P
XML collaboration. However, this proposition replaces deleted elements
by tombstones in the edited document to ensure consistency, making the
document eventually growing without limits and proposes no undo.


Martin et al.~\cite{martin09collaborative} proposes an XML-tree
reconciliation mechanism very similar to a CRDT since concurrent
operations commute without transformation. However, this approach does
not treat XML element's attributes, which require a specific treatment,
since they are unique and unordered. Furthermore, it uses state vector that
limits its scalability and proposes no undo feature.


In the field of Data Management, some works give attention to XML
replication. Some of them \cite{koloniari05peer,abiteboul03dynamic}
suppose the existence of some protocol to ensure consistency of
replicated content without defining it. Finally, \cite{lindholm03xml}
proposes a merging algorithm for concurrent modifications that can only
be used in a centralized context.

\section{XML CRDT without undo}
\label{sec:crdt}

In a collaborative editor, to ensure scalability and
high-responsiveness of local modifications, data must be
replicated. This replication is optimistic since local modifications
are immediately executed. The replicas are allowed to diverge in the
short time, but the system must ensure eventual consistency. When the
system is idle (i.e., all modifications are delivered), the replicas
must have the same content.  

A Commutative Replicated Data Type(CRDT)~\cite{preguica09commutative}
is a data type where all concurrent operations commute. In other
words, whatever the delivery order of operations, the resulting
document is identical. As a result, a CRDT ensure eventual consistency
as proven in~\cite{preguica09commutative}.

Thus, we see an XML collaborative editor as a set of network nodes
that host a set of replicas (up to one per node) of the shared XML
document. Local modifications are immediately executed and
disseminated to all other replicas. We assume that every replica will
eventually receive every modification.

We consider an XML tree as an {\em edge} $e$ with three elements : 
$e.identifier$ the unique identifier of the edge (a timestamp),
$e.children$ the children of the edge (a set of edge),
and $e.attributes$ the attributes of the edge (a map string to
value). The key of the map are the attribute's name (a string), and
a value $av$ has two elements,
$av.value$ : the current value of the attribute (a string),
$av.timestamp$ : the current timestamp of the attribute. The basic
operations that affect an XML tree are :
\begin{itemize}
\item $Add(id_p, id)$ : Adds a edge with identifier $id$ under the
  edge $id_p$. This edge is empty, it has no tag-name, child or
  attribute.
\item $Del(id)$ : Deletes the edge identified by $id$.
\item $SetAttr(id, attr, val, ts)$ : Sets the value $val$ with the
  timestamp $ts$ to the attribute $attr$ of the edge identified by
  $id$. The deletion of an attribute is done by setting is value to
  nil.
\end{itemize}

%
%

To allow $Add$ and $Del$ operations to commute, we use a unique
timestamp identifier. Timestamp identifiers can be defined as follows:
each replica is identified by a unique identifier $s$ and each operation
generated by this site is identified by a clock $h_s$ (logical clock or wall clock).  An
identifier $id$ is a pair $\idComp{h_s}{s}$.  For instance
$\idComp{3}{2}$ identifies the operation $3$ of the site number $2$. The
set of the identifiers is denoted by $\ID$. Thus, two edges added
concurrently at the same place in the tree have different identifiers.
 


To allow $SetAttr$ operations to commute, we use a classical
last-writer-wins technique. We associate to each attribute a timestamp
$ts$. A remote $SetAttr$ is applied if and only if its timestamp is
higher than the timestamp associated to the attribute.
Timestamps are totally ordered. Let $ts_1
= \idComp{h_1}{s_1}$ and $ts_2 = \idComp{h_2}{s_2}$, we have $ts_1 >
ts_2$ if and only if $h_1 > h_2$, or $h_1 = h_2$ and
$s_1>s_2$. Clocks are loosely synchronized, i.e., when a replica
receives an operation with a timestamp $\idComp{h_2}{s_2}$, it sets its
own clock $h_1$ to $max(h_1, h_2)$.
\paragraph{Special attributes.} The special attributes $@tag$ and $@position$ contain the
tag-name and the position of an edge and cannot be nil. 
The position allows to order the children of a node. 
This position is not a basic number. Indeed, to ensure
that the order among edges is the same on all replicas, this position
must be {\em unique, totally ordered and dense}. Positions are dense
if a replica can always generate a position between two arbitrary
positions. This position can be a priority string concatenated with
an identifier~\cite{martin09collaborative}, a sequence of
integers~\cite{weiss09logootundo}, or a bitstring concatenated with
an identifier~\cite{preguica09commutative} all with a lexicographic ordering.
Finally, to model the
textual edges we use another special attribute $@text$. If this
attribute has a value $v$, whatever the value of other attributes, the
edge is considered as a textual edge with content $v$.

%
\paragraph{Algorithms}
The function {\sf deliver($op, t$)} applies an operation $op$ on an XML
tree $t$. The function {\sf find($t, id$)} returns the edge identified
by $id$. The function {\sf findFather($t, id$)} returns the father of
the edge identified by $id$.
\begin{lstlisting}
deliver ($Add(id_p, id), t$) :
  edge $p$ =  find($id_p, t$), $e$ = new edge($id$);
  if $p\neq nil$ then $p.children = p.children \cup \{e\}$;
end 
deliver ($Del(id), t$) :
  edge $p$ =  findFather($id, t$), $e$ =  find($id, p$);
  if $p\neq nil$ then $p.children = p.children \backslash \{e\}$;
end  
deliver ($SetAttr(id, attr, val, ts), t$) :
  edge $e$ =  find($id, t$);
  if $e\neq nil$ and ($e.attributes[attr] = nil$ or  $e.attributes[attr].timestamp < ts$) then 
      $e.attributes[attr].value = val$;
      $e.attributes[attr].timestamp = ts$;
  endif
end  
\end{lstlisting}

\section{XML CRDT with undo}
\label{sec:undo}

Obtaining a correct undo from the user's point of view is a
non-trivial task. Let's have the following scenario 
\begin{inparaenum}
\item a user adds an element,
\item a user deletes this element,
\item the add is undone, 
\item the delete is undone concurrently by 2 different users.
\end{inparaenum}
At the end, since both operations add and delete are undone, the node must be invisible.
And this must be true on every replica and whatever the delivery order of operations.
For instance, using $Del$ to undo $Add$ leads to different results according to the
reception order of the operations. The element is visible if an
un-delete is received in last or not if it is a un-add. Such a behavior
violates eventual consistency.

To obtain a satisfying undo, we keep the information about 
every operations applied to each edge. Then we count the
{\em effect counter} of an operation : one minus the number of undo
plus the number of redo. If this effect counter is greater than 0, the
operation has an effect. An element is visible if the add
has an effect counter greater than 0, and no delete with an effect
counter greater than 0.
%
%
The value of an attribute is
determined by the more recent value with an effect counter greater
than 0. Thus, we need to keep into the map of attributes, the list of
values -- including nil values -- associated to an effect counter. The
list is ordered by the decreasing timestamp.


With undo, an edge attribute $e.attributes[attr]$ becomes an ordered list of {\em
  value} $v$, each value containing 3 elements :
$v.value$ a value of the attribute (a string), 
$v.timestamp$ the timestamp associated to this value,
and $v.effect$ the effect counter of this value (a integer).
{\em The list is ordered by the timestamp}. The function {\sf add($l,
  v$)} adds a value $v$ in the list $l$ at its place according to
$v.timestamp$. The function {\sf get($l, ts$)} returns the value
associated to $ts$ in the list $l$. The special $@add$ attribute has
only one value associated to the timestamp equal to the edge
identifier. The special $@del$ attribute stores the list of timestamp
of delete operations applied to the edge.
\begin{lstlisting}
deliver ($Add(id_p, id), t$) :
  edge $p$ =  find($t, id_p$), $e$ = new edge($id$);
  $p.children = p.children \cup \{e\}$ 
  add($e.attributes[@add]$, new value ($nil, id, 1$));
end  
deliver ($Del(id, ts), t$) :
  edge $e$ =  find($t, id$);
  add($e.attributes[@del]$, new value ($nil, ts, 1$));
end  
deliver ($SetAttr(id, attr, val, ts), t$) :
  edge $e$ =  find($t, id$);
  add($e.attributes[attr]$, new value ($val, ts, 1$));
end  
\end{lstlisting}

Undo of an operation is simply achieved by decrementing the corresponding
effect counter. When a $Redo$ is delivered, the {\sf increment}
function is called with a delta of $+1$.
\begin{lstlisting}
deliver ($Undo(Add(id_p,id)), t$) :
  increment($t, id, @add, id, -1$);
end  
deliver ($Undo(Del(id, ts)), t$) :
  increment($t, id, @del, ts, -1$);
end  
deliver ($Undo(SetAttr(id, attr, val, ts)), t$) :
  increment($t, id, attr, ts, -1$);
end  
function increment($t, id, attr, ts, delta$)
  edge $e$ = find($t, id$);
  value $v$ = get($e.attributes[attr], ts$);
  v.effect += delta;
end
\end{lstlisting}

\paragraph{Exemple.}

Figure~\ref{fig:concurUndosDegree} presents the application of our
functions on the introducing example.
On every replica, the add and del operations have an effect
counter lesser or equal to 0. 
Thus none of these operations have an effect on the XML tree and the edge is invisible.

\newcommand{\ei}{*+[F]{\txt{$(id_p,[\ldots],\{\})$}}}
\newcommand{\ea}{*+[F]{\txt{$(id_p,[\ldots],$ \\ $\{(id,\{\}, [@add
      \map (nil, id, 1)])\})$}}}
\newcommand{\ed}{*+[F]{\txt{$(id_p,[\ldots], \{(id,\{\}, [@add
      \map (nil, id, 1),$ \\$ @del \map (nil, ts, 1)])\})$}}}
\newcommand{\eud}{*+[F]{\txt{$(id_p,[\ldots], \{(id,\{\}, [@add
      \map (nil, id, 1),$ \\$ @del \map (nil, ts, 0)])\})$}}}
\newcommand{\euda}{*+[F]{\txt{$(id_p,[\ldots], \{(id,\{\}, [@add
      \map (nil, id, 0),$ \\$ @del \map (nil, ts, 0)])\})$}}}
\newcommand{\eudd}{*+[F]{\txt{$(id_p,[\ldots], \{(id,\{\}, [@add
      \map (nil, id, 1),$ \\$ @del \map (nil, ts, -1)])\})$}}}
\newcommand{\eudda}{*+[F]{\txt{$(id_p,[\ldots], \{(id,\{\}, [@add
      \map (nil, id, 0),$ \\$ @del \map (nil, ts, -1)])\})$}}}

\begin{figure}[htb]
  \centerline{ \scalebox{0.75}{ \xymatrix@C=15pt@M=2pt@R=15pt{
        *+[F-,]\txt{Replica 1} \ar@{.}'[d]'[dd]'[ddd]'[dddd]'[ddddd]'[dddddd]'[ddddddd]'[dddddddd]'[dddddddddd][ddddddddddd] & *+[F-,]\txt{Replica 2} \ar@{.}'[d]'[dd]'[ddd]'[dddd]'[ddddd]'[dddddd]'[ddddddd]'[dddddddd]'[dddddddddd][ddddddddddd]& *+[F-,]\txt{Replica 3} \ar@{.}'[d]'[dd]'[ddd]'[dddd]
\\
   \ei & \ei & \ei \\
   Add(id_p,id)        &   Add(id_p, id) \ar[r] \ar[l]  &    Add(id_p,id)                                \\
   \ea & \ea & \ea \\
   Del(id, ts)           &   Del(id, ts) \ar[l]           &    undo(Add(id_p,id)) \ar@/_0.5pc/[ddddll] \ar@/^5pc/[ddddddl] \\
   \ed & \ed  \\
   undo(Del(id, ts))\ar[ddr]     &   undo(Del(id, ts)) \ar@/_2pc/[ddddl]         &                                    \\
   \eud & \eud   \\
   undo(Add(id_p,id))     &  undo(Del(id, ts))           &                                         \\
   \euda & \eudd   \\
   undo(Del(id, ts))     &   undo(Add(id_p,id))          &              \\
   \eudda & \eudda   \\
        }} }
    \caption{Concurrent undos with effect counters.}
  \label{fig:concurUndosDegree}
\end{figure}

\paragraph{Model to XML.}

As the model described above includes tombstones and operations
information, it cannot be used directly by applications. Indeed,
applications must not see a tombstones and only one value for each 
attribute.  A node is
visible if the effect counter of the attribute $@add$ is at least
one, and if all values of the attribute $@del$ have an effect
counter of at most 0. If the edge is a text, i.e.,
the attribute $@text$ has a value, we write this value in the XML
Document.
If the edge is visible and is not a text, we write the tag
and the attributes corresponding to that edge.  Therefore, we need to compute the current value of the
attributes which is the newest non-undone value of a value list.
Finally, the rendering function calls itself to treat the children of the
edge.

\paragraph{Correctness.}

To ensure that our data type is a CRDT and thus that eventual
consistency is ensured, we must prove all our operations commutes. For
a complete proof, please see~\cite{martin10scalableRR}. The
only requirement to ensure consistency of the XML CRDT without undo is
to receive delete operation after insert of a node. With undo, this
constraint is not required to ensure consistency since a delete can be
received before an insert. The delete produces directly a tombstone.

\section{Garbage collecting}
\label{sec:scalability}

Concerning the scalability in term of operations number, the
XML CRDT without undo requires tombstones for attributes as the Thomas
Write Rule defined in the RFC~677~\cite{RFC0677}. 
The XML CRDT with undo requires to keep an information about every 
operation applied to the XML document.
This is not surprising since any undo system must keep a trace of an operation 
that can be undone, either in the document model or in a history log.

%
%
%
However, a garbage collecting mechanism can be designed. Such a
garbage collection is similar to the one already present in the
RFC~667~\cite{RFC0677}.  Each replica $i$ maintains a vector $v_i$ of
the last clock timestamp received by all other replicas (including its
own clock).  From this vector the replica computes $m_i$ the minimum of these
clocks. This minimum is sent regularly to the other replicas. It can
be piggybacked to operation's messages or sent regularly in a specific
message. From the minimum received (including its own), each replica
maintains another vector $V_i$. The minimum of $V_i$ is $M_i$. The
point is that, if communication is FIFO, a replica knows that every
replica has received all potential messages with a timestamp less or
equal to $M_i$. Thus any tombstone with a timestamp less or equal to
$M_i$ can be safely removed. This mechanism can be directly used in the
XML CRDT without undo to remove old deleted attributes.

In the XML CRDT with undo, we only authorize to produce an undo of an
operation whose timestamp is greater than $m_i$. Thus operations with
a timestamp lesser than $M_i$ will never see their effect
modified. So, elements such as follows can be safely and definitively
purged :
\begin{itemize} 
\item attribute value $v$ with $v.timestamp < M_i$ and $v.effect \leq 0$
\item attribute value $v$ with $v.timestamp < M_i$ and there exists 
  $v'$ with $v.timestamp < v'.timestamp < M_i$ and $v'.effect > 0$
\item attribute with no value or with every value $v$ such that 
   $v.timestamp < M_i$ and ($v.effect \leq 0$ or $v.value = nil$)
\item edge with any delete value $d$ with $d.timestamp < M_i$ and
  $d.effect > 0$ or with the add value $a$ with $a.timestamp < M_i$
  and $a.effect \leq 0$.
\end{itemize}
Thus, the time and space complexity of the approach is greatly reduced
to be proportional to the size of the view. Moreover, differently to
the RFC 677, replicas send $m_i-k$ with $k$ a global constant
instead of $m_i$. Thus, even if the replicas are tightly synchronized --
having $m_i$ very close to their own clock --, the replicas can always
undo the last operations.
Also, the garbage collecting mechanism that can
be adapted to the other tombstone-based approach 
is much less scalable since based on a
consensus-like method~\cite{letia09crdt}.





\section{Conclusion}
\label{sec:conclusion}

We have presented a commutative replicated data type that supports XML
collaborative editing, including a global selective undo mechanism. Our
commutative replicated data type is designed to scale since the
replicas number never impacts the 
execution complexity. Obviously, the undo mechanism requires to keep information
about the operations we allow to undo. We presented a garbage
collection mechanism that allows to purge the old operations
information. 

We still have much work to achieve on this topic. Firstly, we need to
make experiments to establish the actual scalability and efficiency of
the approach in presence of huge data. Secondly, we plan to study
replication of XML data typed with DTD or XSD. This is a difficult task,
never achieved in a scalable way.

\bibliographystyle{abbrv}
\bibliography{theBib}



\end{document}